\documentclass[12pt,preprint]{aastex}
\usepackage{psfig}
\usepackage{natbib}

\shorttitle{Is BL Lac Precessing?}
\shortauthors{Mutel \& Denn}

\def\BL{BL Lac\ }

\def\edcomment#1{\iffalse\marginpar{\raggedright\sl#1\/}\else\relax\fi}
\reversemarginpar

\received{2004 May 19}
\begin{document}
\title{Is the Radio Core of \BL Precessing?}
\author{R. L. Mutel}
\affil{Dept. Physics \& Astronomy, University of Iowa, Iowa City IA
52242}
\email{robert-mutel@uiowa.edu}
\author{G. R. Denn}
\affil{Sweet Briar College, Sweet Briar, VA 24595}
\email{gdenn@sbc.edu}

\begin{abstract}
\citet{s03} reported the discovery of a 2.3 year periodic variation in
the structural position angle of the parsec-scale radio core in the
blazar \BL. We searched for independent confirmation of this periodic
behavior using 43~GHz images of the radio core during ten epochs
overlapping those of Stirling et al. Our maps are consistent
with  several periodicities, including one near the period reported by
Stirling et al. By comparing our position angle
measurements with those of Stirling et al., we find strong consistent
evidence for position angle variations of the inner core during the
observed epochs.  However, the claim for periodic variation is not
convincing, especially when the most recent epochs (2000.60 - 2003.78) are
included. A definitive resolution will require continued monitoring of
the core structure over several periods.  \end{abstract}


\section{Introduction}
There is growing evidence that at least some active galactic nuclei (AGN)
display periodic behavior on a timescale of years or decades. The
observational evidence
consists largely of periodicities in  optical or radio light curves or
periodic variations in the parsec-scale radio cores mapped by VLBI.
Perhaps the best documented case study is that of the BL Lac object OJ287
for which a period of 11.6 year was found in the visible light curve
\citep{s88}. This periodicity was confirmed by observations of a
predicted outburst in 1994 \citep{s96,p00} but with a one year delay between
the optical and radio peaks \citep{v99}. \citet{k97} suggested that the
observed periodicity is due to the sweeping of a relativistic beam
aligned normal to a precessing accretion disk surrounding a supermassive
black hole. \citet{v00} suggest a binary black hole model in which the
secondary orbit penetrates the accretion disk of the primary, causing a
thermal (optical) pulse, followed by enhanced accretion and increased
particle flux in the relativistic jet, which is responsible for the
time-delayed radio flare. 

Precessing relativistic beams have also been
invoked to explain the undulating jet structures in several compact radio
sources, including 1928+738 \citep{r93,mph03}, 0153+744 \citep{h97}, and
4C12.50 \citep{L03}.  \citet{b00,b01} interpreted the strongly bent
relativistic jet in PKS 0420-014 as evidence for a binary black hole
system which precesses on a timescale $\sim$~10 yr.  \citet{a99} analyzed
eight superluminal jet components of the  parsec-scale radio quasar 3C273
observed over the last 30 yr. They suggest that the observed velocities
and position angles can be interpreted by ejection of
synchrotron-radiating shocks moving outward in ballistic trajectories
from the base of a precessing inner jet whose period is 16 yr. 

These observations provide {\it post facto} evidence for precession in
the central engines of AGN i.e., the observed jet morphology is
interpreted as a record of the putative jet motion, but the jet
oscillation itself has not been observed. However, in a recent paper
\citet{s03} report that they have directly detected periodic changes in
the parsec-scale radio jet orientation of the eponymous AGN BL Lacertae
based on  analysis of two
independent datasets: Periodic variations in the polarization position
angle at 1 mm wavelength, and in the direction of the
innermost radio core component in 43~GHz VLBI maps.  In the latter
case, they find strong evidence of periodicity with a timescale of
$2.29\pm0.35$ yrs and an angular amplitude of $24.4\degr\pm16\degr$. The
VLBI maps spanned 23 epochs over the time range 1997.58 -- 2001.28. They
modeled the source brightness at each epoch using a variable number of
elliptical Gaussian components, including a two closely-spaced ($\sim0.1$
mas) circular Gaussian subcomponents in the core. The model component
parameters were adjusted for a best-fit to the observed brightness
distribution. They found a periodic variation in the core `structural
position angle' (SPA), defined as the relative position angle between the
two closely spaced core subcomponents.

This paper reports results from a series of 43~GHz VLBI
observations of BL Lac using very similar observing parameters during an
overlapping time range (1998.73 -- 2003.82). Although the primary purpose
of the observations was rather different (monitoring the magnetic field
structure of the radio jet, Mutel \& Denn, 2004 {\it in prep.}), the
resulting maps are of comparable quality to those of \citet{s03} and
provide an independent test of the precession hypothesis.

\section{Observations and Data Reduction} The observations were performed
at regular $\sim$0.3 year intervals during nine epochs between 1998.76
and 2002.05 and a tenth epoch at 2003.82 using the 10-element Very Long
Baseline Array
\footnote{The National Radio Astronomy
Observatory is operated by the National Radio Astronomy Observatory by
Associated Universities, Inc., under cooperative agreement with the
National Science Foundation} (VLBA).  The observations consisted of a
series of short scans at each of three observing frequencies (15.4, 22.2,
43.2~GHz) of BL Lac and three other AGN sources in sequence for 12 hours.
We recorded both left and right circular polarization at each telescope
and correlated all four cross-correlations pairs (RR, LL, RL, and LR) for
each baseline. The data were recorded in standard VLBA mode using 128~MHz
bandwidth and 1-bit sampling. 

Prior to self-calibration, all fringe-fitting and visibility calibration
was done using the Astronomical Imaging Processing System 
\citep[AIPS,][]{vkg96}.  We performed amplitude calibration using the AIPS
automated calibration transfer system \citep{u99}. Fringe fitting of the
parallel (RR, LL) and cross-hand (RL, LR) polarization correlations, as
well as removal of instrumental polarization were done using
the scheme described in \citep{dmm00}. Hybrid mapping was
done in AIPS and with the Caltech hybrid mapping program DIFMAP
\citep{s97}. The resulting Stokes I maps had typical RMS noise
level $\sim0.8$ mJy per beam at 43~GHz. 

\clearpage
\begin{figure*}
\epsscale{1.0}
\plotone{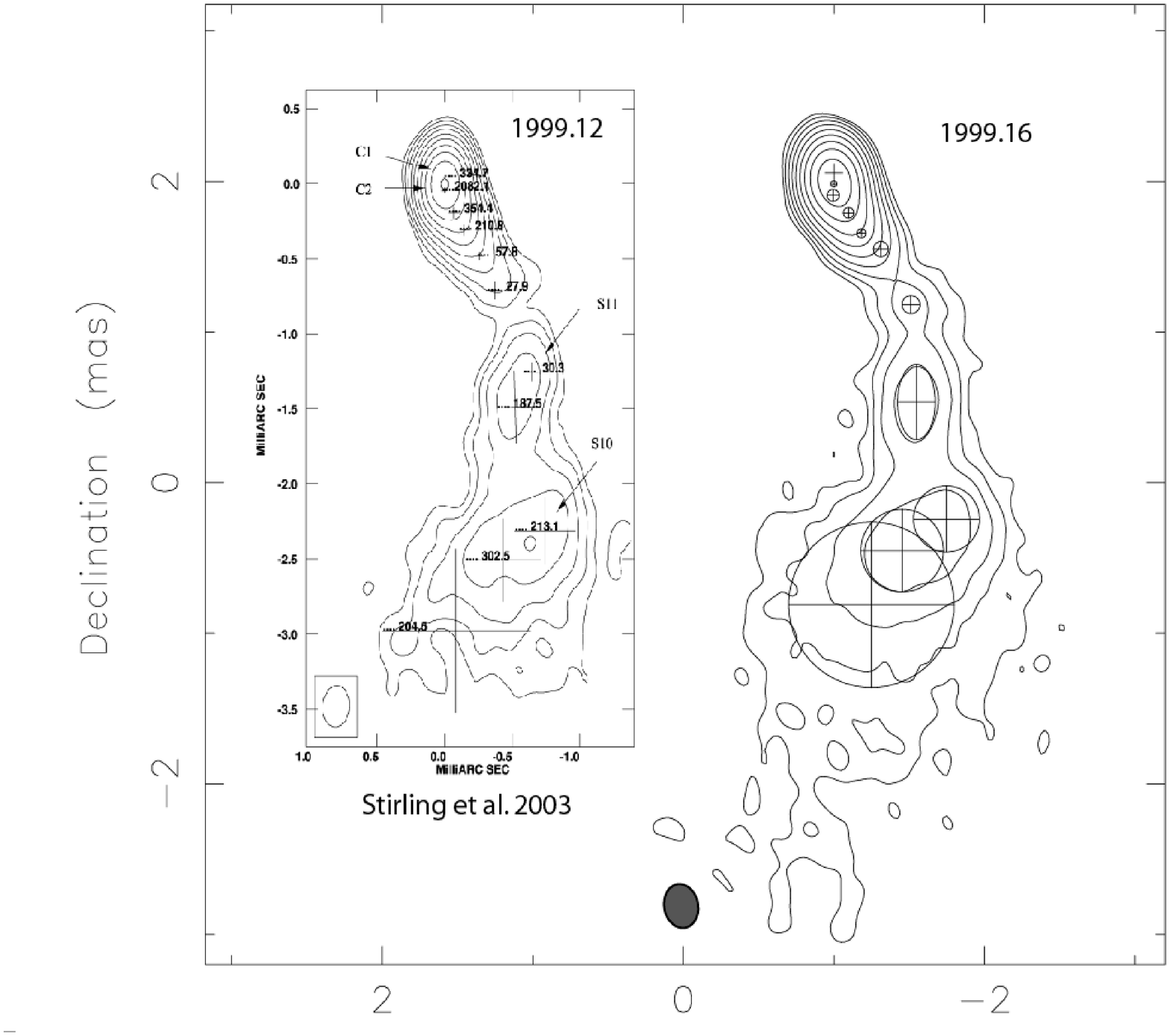}
\caption{43~GHz contour maps of the BL Lac's radio core at epochs
1999.12 \citep[left inset, ][]{s03}  and 1999.16 (right, present data). Contour
levels increment in factors of two starting at 3 mJy in both maps. The
overlaid crosses and  circles with crosses indicates the size of
fitted elliptical Gaussian components.} \end{figure*}
\clearpage

Figure 1 shows a comparison of 43~GHz  contour maps of BL Lac observed by
\citep{s03} at epoch 1999.12 and our map 14 days later (epoch 1999.16).
In both maps, eleven component best-fit elliptical Gaussian brightness 
models are shown overlaid by crosses and circles.
The maps are very similar, both in dynamic range and angular resolution,
as might be expected given the similarity in observing equipment and
close proximity in time. However, there are small but statistically significant
differences in the Gaussian component locations and sizes, which 
might reflect real differences in structure over 14 days, but could also
be partially explained by subtle differences in the model fitting
algorithms used-- \cite[Brandeis VLBP package,][]{r87} versus
\cite[DIFMAP,][]{s97}.  Nevertheless, the close similarity in image
quality and resolution of the two datasets confirms that the present data
comprise a comparable but independent test of the claim for periodic
behavior in the inner core. 

Gaussian component modeling of the inner core structure is illustrated in
Figure 2. Figure 2$a$ shows the full 43~GHz hybrid map of BL~Lac at epoch
1999.73, while Figure 2$b$ is a magnified view of the core only. The
innermost part of the core was fit with three circular Gaussian
components labeled C1, C2, and C3. The structural position angle (SPA) is
defined as the position angle between the two innermost components C1 and
C2. It is this orientation angle that \citet{s03} find exhibits periodic
variation which they interpret as evidence for core precession. Note that
the C1-C2 separation is only $\sim20$\% of the restoring beam
dimension along the structural position angle (lower
left ellipse). Hence, the inner core substructure is not well resolved,
which makes the Gaussian component decomposition non-unique. To check the
reliability of the SPA determination, both \citet{s03} and us also fitted
a single elliptical Gaussian component to the innermost core for each
epoch. The resulting major axis orientation is compared with the SPA
orientation using the C1-C2 position angle in Table 1. The differences
are less than the uncertainties in the position angle determination,
indicating that although the angular resolution of the 43~GHz maps is
inadequate to determine details of the inner core substructure, the
orientation of the inner core is probably accurately determined by
Gaussian component modeling.

\clearpage
\begin{center}
\begin{deluxetable}{l l l l r}
\tablecaption{Gaussian Model Parameters}
\tablehead{
\multicolumn{1}{c}{Epoch} &  
\multicolumn{2}{c}{\underline{2 Component Gaussian}} &
\multicolumn{1}{c}{ \underline{Ellip. Gaussian}}  & 
\multicolumn{1}{c}{$\Delta$SPA} \\
\multicolumn{1}{c}{yr}  & 
\multicolumn{1}{c}{SPA($\degr$)} &  
\multicolumn{1}{c}{r(mas)} & 
\multicolumn{1}{c}{SPA($\degr$)}  &  
\multicolumn{1}{c}{$\Delta$ ($\degr$)} 
}
\tablewidth{0pt}
\startdata
1998.76 & $207.2\pm9.3$ & $0.07\pm0.014$ & $200.0\pm10.6$ & $7.2\pm14.1$ \\
1998.97 & $191.5\pm8.3$ & $0.08\pm0.009$ & $190.1\pm7.1$ & $1.4\pm10.9$ \\
1999.16 & $184.3\pm6.8$ & $0.08\pm0.007$ & $180.8\pm7.4$ & $3.5\pm10.0$ \\
1999.41 & $193.0\pm8.5$ & $0.07\pm0.009$ & $194.3\pm9.2$ & $-1.3\pm12.5$ \\
1999.73 & $180.3\pm9.1$ & $0.06\pm0.007$ & $170.9\pm7.9$ & $9.3\pm12.1$  \\
2000.01 & $176.7\pm8.9$ & $0.09\pm0.007$ & $179.0\pm8.5$ & $-2.3\pm12.2$ \\
2000.31 & $194.3\pm4.5$ & $0.12\pm0.007$ & $194.7\pm4.3$ & $-0.4\pm6.2$   \\
2001.60 & $188.9\pm5.3$ & $0.14\pm0.008$ & $185.2\pm5.2$ & $3.7\pm7.5$ \\
2002.05 & $186.9\pm6.8$ & $0.15\pm0.008$ & $187.5\pm6.9$ & $-0.6\pm9.7$   \\
2003.82 & $172.3\pm5.8$ & $0.09\pm0.008$ & $175.0\pm8.4$ & $-2.7\pm10.2$ \\ 
\enddata
\end{deluxetable}
\end{center}
\clearpage

\subsection{Model Component Uncertainty Estimates}
We used the program  DIFWRAP \citep{L00} to estimate uncertainties in
fitted Gaussian components. DIFWRAP is a `front-end' graphical user
interface to DIFMAP which allows the user to constrain a selected
subset of model parameters over a fixed grid of values while allowing
other parameters to freely vary. For each set of model parameters, the
visibilities are self-calibrated and hybrid mapped, with the adjustable
model components varied so ensure a best-fit to the visibility data. This
method has the advantage that it allows for the inter-dependence of model
components to be taken into account during model fitting. The resulting
self-calibrated maps are displayed for visual inspection with the
$\chi^2$ value of each set of model components.
  
We used DIFWRAP to vary the separation and SPA of the core subcomponents
C1 and C2 in steps of 0.02 mas and 5$\degr$ respectively, followed by
self-calibration and model fitting for each set of C1-C2 positions.
We generated a sequence of hybrid maps corresponding to gridded values of
separation and SPA over the ranges 0.02 to 0.20 mas and 180$\degr$ to
240$\degr$ respectively. At each epoch, 
the map corresponding to the minimum
$\chi^2$ value was differenced from each of the other maps. 
In order to determine the uncertainties in the separation and SPA of
the C1-C2 component pair, 
we adopted the heuristic criterion that the RMS noise
level of the differenced map associated with a particular core separation and
SPA did not increase by more than 50\% from the lowest value for any map.  

Comparison of the SPA uncertainties of \citet{s03} with the present data
shows that the former are significantly smaller than ours: The mean
uncertainties are $\pm5.0\degr$ and $\pm7.3\degr$ respectively. This is
surprising, since although the detailed method to determine
uncertainties differs, the instrumental technique, mapping procedures,
and resulting maps are nearly identical (e.g. Figure 1). More important,
the published \citeauthor{s03} SPA data does not agree
with their proposed periodic model: The formal fit of SPA angle vs. epoch
to the periodic function described in \citet{s03} can be rejected
at 99.9\%
confidence. Hence, in order to compare our SPA data with those of
\citet{s03} in a consistent manner, we solved for a constant
multiplicative factor to apply to their SPA uncertainties which would
result in a reduced $\chi^2$ = 1 when their model was applied. The
resulting factor (1.5) made their uncertainties comparable with ours. In
the following, we refer to the Stirling et al. dataset with larger
uncertainties as the rescaled Stirling et al. data set.

\clearpage
\begin{figure} 
\epsscale{1.0}
\plotone{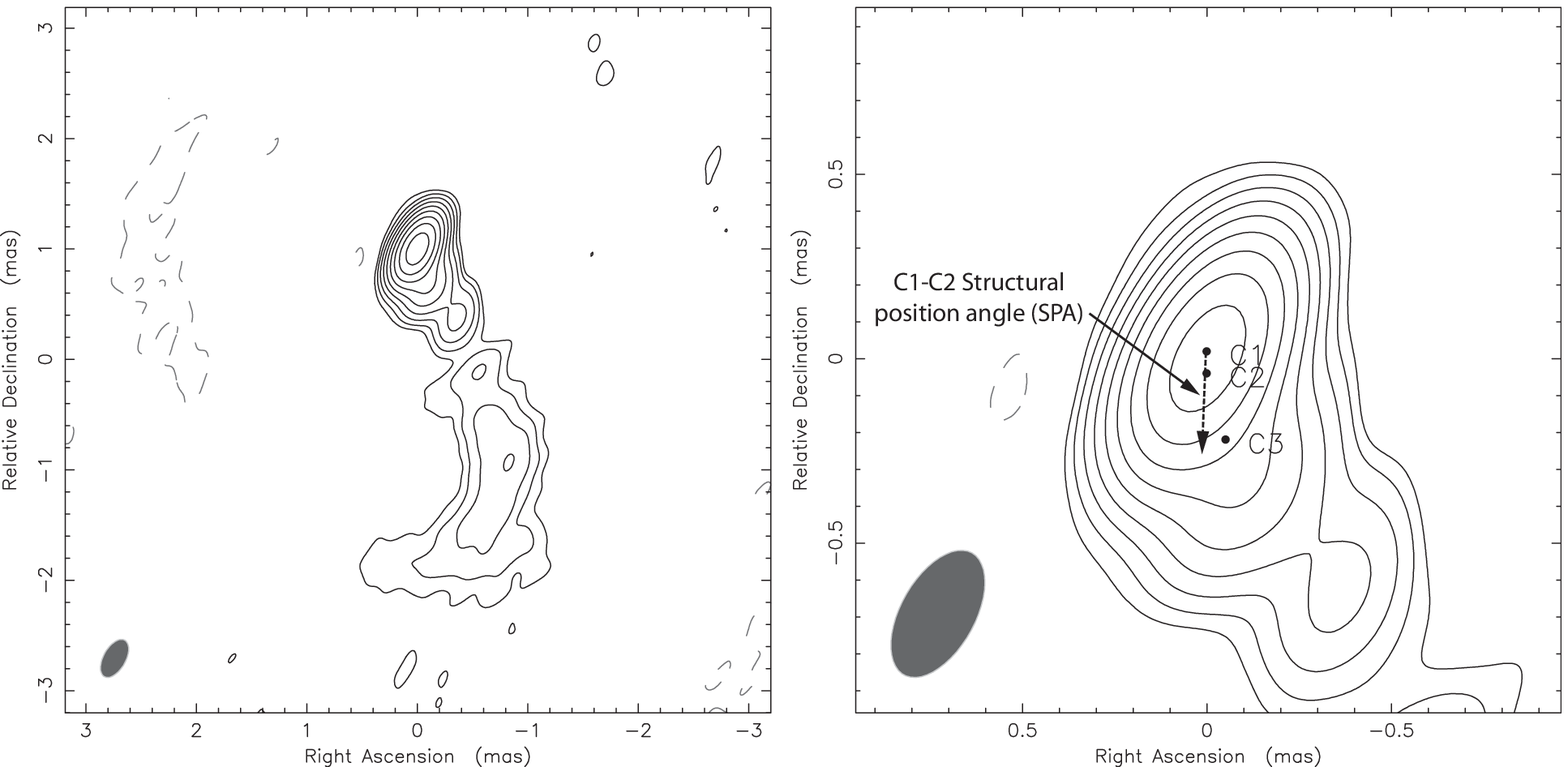} 
\caption{\scriptsize{$(a)$ 43~GHz map of BL Lac at epoch 1999.73. Contour
levels increase in factors of 2 from 3.77 mJy per beam with a peak of
1510 mJy per beam. Dashed contours are negative. 
$(b)$ Same as $(a)$, but expanded to show only the inner core and
centroids of 
best-fit Gaussian subcomponents C1, C2, and C3. The structural
position (SPA) used to determine the orientation of the inner core is
defined as the angle between subcomponents C1 and C2. For this epoch, the
SPA was $180.3\degr\pm11.7\degr$}} 
\end{figure}

\section{Results} The variation of inner core SPA's versus epoch were
compared with the periodic model of \citeauthor{s03} as well as
least-squares fitted periodic and constant models. We analyzed 
measurements derived from the ten epochs reported in this paper, the 
data of \citeauthor{s03}, and a combined dataset consisting of all 33
epochs. The fitted parameters as well as resulting reduced $\chi^2$ and
agreement factors are listed in Table 2. The periodic models 
have the form
\begin{equation} \phi_{spa} (t) = A_0 \sin \left( {2\pi  \cdot
\frac{{\left( {t - t_{_0 } } \right)}}{P}} \right) + \phi _0
\end{equation}  
Column 1 lists the model being tested, while columns 2-5 list the model
parameters. Columns 6 and 7 list the SPA dataset and number of epochs.
Column 8 lists the reduced $\chi^2$ of the model fitted to the data,
while column 9 lists the corresponding $\chi^2$ goodness of fit probability 
(i.e., the
probability that an experimental data set consistent
with the assumed model would result in a reduced $\chi^2$ value at least
as large as the measured value).  We refer to this as the acceptance
probability in the following.

Figure 3$a$ is a plot of core SPA (determined from C1-C2 position angle)
versus epoch for all ten epochs listed in Table 1, along with the null
hypothesis model (non-varying SPA, solid line). The reduced $\chi^2$ of
the fit is 2.76 (acceptance probability $p=0.01$). Hence, at 99\%
($3\sigma$) confidence 
the present SPA data indicates significant variations in the
orientation of the inner core. Similarly, the one component
elliptical Gaussian SPA data has a reduced $\chi^2$ of 2.24
($p=0.04$), implying significant variability at $2\sigma$
confidence. The \citeauthor{s03} SPA data also indicate variability at
high confidence ($p<10^{-4}$). 

Figure 3$b$ shows the same data, but with a least-squares fit to a
periodic model (solid line, period 13.1 yrs), and the
model of \citeauthor{s03} (dotted line). While neither model can
be excluded, the fits are not
particularly good: The acceptance probabilities are $p=
0.14, 0.10$ respectively. By comparison, the  one component
elliptical Gaussian SPA data has a best-fit
solution with a 1.43 yr period and is in good agreement
($p=0.38$). The large difference in least-squares periodic
solutions between the one and two-component SPA data, in spite of
their similar values (Table~1), indicates that ten sampling epochs
is probably insufficient to determine the true periodicity of the
SPA data if it exists.

Figures 3$(c)$ shows combined SPA measurements from
\citeauthor{s03} (open circles) along with the present data
(solid squares), and the \citeauthor{s03} periodic model (dashed
line). The reduced $\chi^2$ of the fit is 1.88, or an acceptance
probability $p=0.003$. However, we have argued (see \S2.1) that
the uncertainties reported by \citeauthor{s03} may be
underestimated. Figure 3$(d)$ shows the same data, but with
rescaled \citeauthor{s03} uncertainties.  The \citeauthor{s03}
model (dashed line) and the least-squares best-fit periodic model
(solid line) are also shown. The corresponding acceptance
probabilities are 0.40 and 0.49 respectively. The best-fit period
is $2.24\pm0.20$ yrs, in excellent agreement with the $2.29\pm0.35$ yr
period determined by \citet{s03}. 

\clearpage
\begin{deluxetable}{l c c c c l c c c }

\tablecaption{
Model fits to BL Lac core structural position angles versus epoch.
}

\tablehead{
Model  & \multicolumn{4}{c}{\underline{Model Parameters}} & 
SPA Data & Epochs  & ${\chi_r}^2$ & Prob. \\
 & $A_0$ & $P(yr)$ & $t_0$ & $\phi_0$ & & & & 
}

\tablewidth{0pt}
\startdata
Stirling et al.&12.2 & 2.29 & 1997.9 & 191$\degr$ & This paper C1-C2 & 10
& 1.79 & 0.10 \\
 & " & " & " & " & This paper Gaussian     & 10 & 1.47 & 0.18 \\
 & " & " & " & " & Stirling et al.         & 23 & 2.28 & 0.001 \\
 & " & " & " & " & Stirling rescaled       & 23 & 1.01 & 0.44 \\
 & " & " & " & " & Combined                & 33 & 1.88 & 0.003 \\
 & " & " & " & " & Combined rescaled       & 33 & 1.05 & 0.40 \\
\hline
Best-fit periodic &12.1&13.1&1996.3&179$\degr$ & This paper C1-C2& 10 &
1.61 & 0.14 \\
                  &12.0&1.43&1997.2&183$\degr$ & This paper Gaussian& 10 & 1.07 & 0.38 \\
                  &12.2&2.29  &1997.9&191$\degr$   & Stirling et al.   & 23 & 2.27 & 0.001 \\
                  &12.2&2.29 &1997.9&191$\degr$   & Stirling rescaled  & 23 & 1.01 & 0.45 \\
                  &12.1&2.24 &1997.9&191$\degr$   & Combined          & 33 & 1.82 & 0.004 \\
                  &12.1&2.22 &1997.9&191$\degr$   & Combined rescaled & 33 & 0.99 & 0.49 \\
\hline
Non-varying & - & - & - & 187$\degr$ & This paper C1-C2 & 10 & 2.78 & 0.01 \\
 & - & - & - & 192$\degr$ & This paper Gaussian   & 10 & 2.24 & 0.04 \\
 & - & - & - & 192$\degr$ & Stirling et al.   & 23 & 7.12 & $<10^{-4}$ \\
 & - & - & - & 192$\degr$  & Stirling rescaled & 23 & 3.16 & $<10^{-4}$ \\
 & - & - & - & 191$\degr$  & Combined          & 33 & 5.35 & $<10^{-4}$ \\
 & - & - & - & 192$\degr$  & Combined rescaled & 33 & 4.98 & $<10^{-4}$ \\
\hline
\enddata
,\end{deluxetable}
\clearpage

\clearpage
\begin{figure} 
\epsscale{1.0}
\plotone{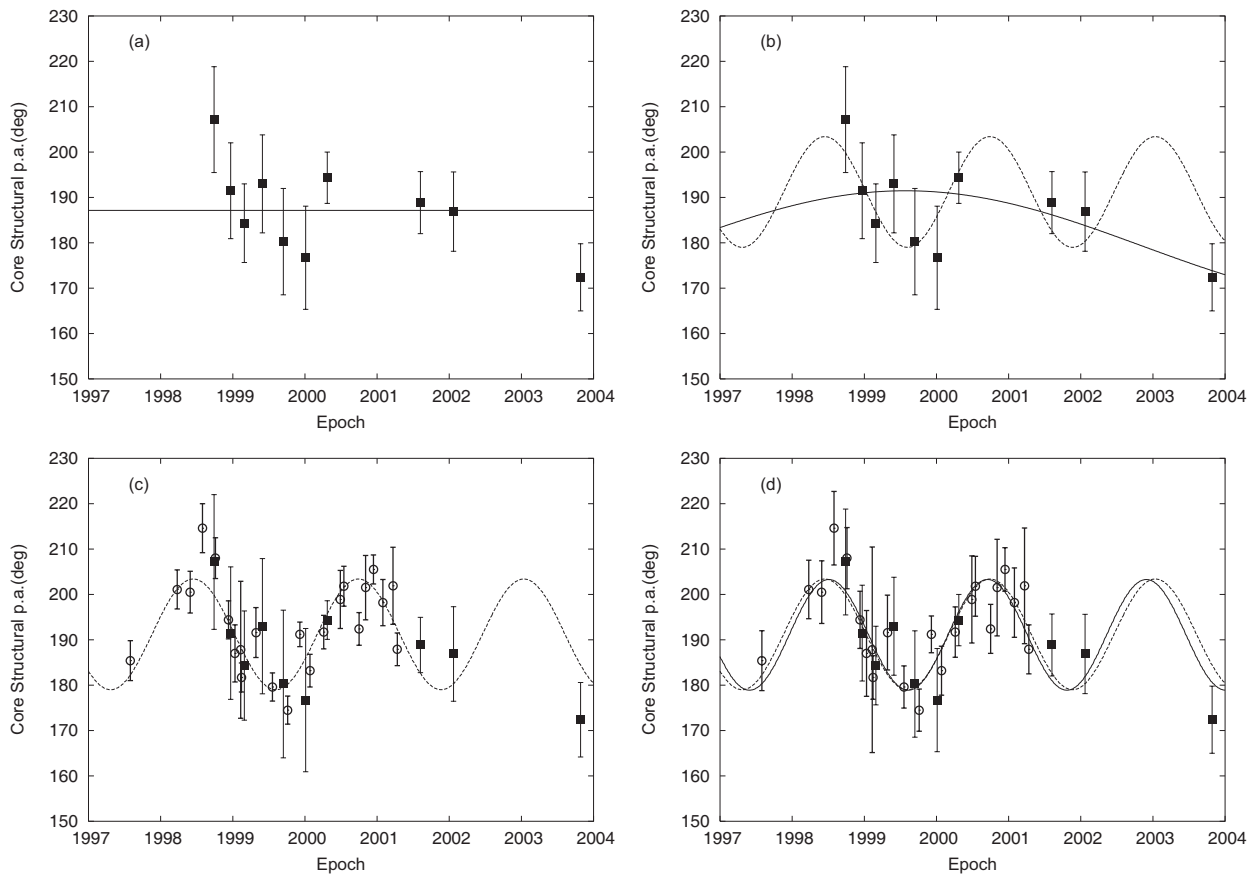} 
\caption{\scriptsize{$(a)$ BL Lac inner core structural position angle
(SPA) vs. epoch for present observations (filled squares) and best-fit
null hypothesis (non-varying) model (solid line).  $(b)$ Same as $(a)$,
but for best-fit periodic model (solid line) and original Stirling et al.
(2003) periodic model (dashed line), $(c)$ Combined SPA data of this
paper (filled squares) and those of Stirling et al. (open circles) and
the periodic model of Stirling et al. (dashed line). $(d)$ Same as $(c)$,
but Stirling et al. SPA uncertainties have been scaled by a constant
factor x1.5 (see text). The periodic model of Stirling et al.  (dashed
line) and best-fit periodic function (solid line) are superposed.}}
\end{figure} \clearpage

\section{Discussion}
Do the present observations support Stirling et al.'s claim of
core orientation periodicity? The results are ambiguous: While
the SPA's derived from the present data are consistent with the
Stirling et al. model, they are also consistent with several
other periodic models. This is not surprising given the
relatively large uncertainty of each SPA measurement with respect
to the half-amplitude of the putative periodic variations.  

Given this ambiguity, it is useful to inquire
whether agreement with the \citeauthor{s03} model is statistically
significant. The acceptance probability of the combined dataset to the
best-fit periodic model ($p=0.49$) is only marginally better than
that of the rescaled \citeauthor{s03} dataset alone ($p=0.45$).
However, even with the relatively large uncertainties of the
present SPA data, would a different set of measurements been able
to falsify the periodic model of \citeauthor{s03}?  In order to
test this, we randomly reassigned the epochs of our ten SPA
measurements and  calculated the goodness of fit to the
\citeauthor{s03} model. We generated 10,000 shuffled datasets and
calculated the resulting reduced $\chi^2$ and acceptance
probabilities for all cases. More than 96\% of these shuffled
datasets had acceptance probabilities $p<0.05$, i.e., they could
be rejected at 95\% ($2\sigma$) confidence or higher. We conclude
that the current SPA data does provide independent confirmation
(at $2\sigma$ confidence) of SPA variations similar to those
reported by \citeauthor{s03} during the time interval spanned by
both sets of observations.

A more critical question is whether the combined dataset provides
convincing evidence for (or against) periodic SPA variations. Since the
combined dataset spans less than two periods with
adequate sampling intervals
(1998.23 - 2002.05), it is difficult to to make  strong case for
periodicity as opposed to episodic variations. If the variations are
regular and periodic, measurement of SPA orientations over a longer
timescale provides a
critical test of periodicity.  We
can use the three epochs observed since the last SPA measurement of
\citeauthor{s03} (epochs 2001.60, 2002.05, 2003.82) to test periodic
models.  A $\chi^2$ goodness of fit analysis of these three measurements
using the periodic model of \citeauthor{s03} and the best-fit model in
Table~2 results in acceptance probabilities $p=0.08$
i.e., rejection of the model at slightly less than $2\sigma$
confidence. While this result is not definitive, it calls into
doubt the periodic nature of the SPA variations.

If the inner core of BL~Lac is precessing, one might expect
periodic variations in the radio flux, since the relativistically
beamed  jet emission
will be strongly modulated by the line
of sight angle. The Doppler-boosted flux will depend on the
Lorentz factor of the emission region and its angle to the
observer's line of sight.  \citet{dmm00}, in  analyzing 17 epochs
of BL~Lac radio jet maps, fit a helical jet model
inclined at an a angle of $9\degr$ to the observer's line of sight.
The four observed superluminal component Lorentz factors were 
in the range $2.8<\gamma<7.0$ ($H_0 = 70$ km s$^{-1}$ Mpc$^{-1}$). 
Since the observed half-angle of precession is $12\degr$, the
deprojected angle is $\pm1.8\degr$. The expected flux ratio
would be 
\[
\frac{{S_{\max } }}{{S_{\min } }} = \left[ {\frac{{\delta \left(
{\gamma ,\theta _{\max } } \right)}}{{\delta \left( {\gamma
{\rm{,}}\theta _{{\rm{min}}} } \right)}}} \right]^{2 + \alpha } 
\]
where $\delta$ is the Doppler factor, $\alpha$ is the source
spectral index,  and $\theta_{max,min} = 9\degr\pm1.8\degr$ are
the extremal values of the line of sight angle. At centimeter
wavelengths BL~Lac is nearly always a flat spectrum source
($\alpha\sim0$),
so the expected flux ratio
over one precession cycle ranges from 1.1 ($\gamma=2.8$) to 2.5
($\gamma=7.0$). \citet{s03} used their kinematic model for BL~Lac
and obtained similar results.

\citet{k03} have recently reported detection of
quasi-periodic variations in the radio flux of BL~Lac. They
analyzed 20 years of synoptic radio flux measurements at 4.8,
8.4, and 14.5~GHz from the University of Michigan Radio Astronomy
Observatory (UMRAO) using a cross-wavelet transform algorithm.
They report significant frequency-dependent peaks: 1.4 yr
(4.8~GHz), 3.7 yr (8.4~GHz) and multiple periods (0.7, 1.6, and
3.5 yr) at 14.5~GHz. These are near, but not definitely not within
the uncertainty range of the proposed precession period of
$2.29\pm0.35$ yr of \citeauthor{s03}, or the amended period
$2.24\pm0.20$ yr determined using all 33 epochs. 
\citet{v04} have searched for
periodicities in  historical radio
and optical light curves for BL~Lac from 1968-2003. They analyzed
multi-frequency radio observations from 4.6 to 37~GHz using
several statistical methods suitable for unevenly spaced samples.
Although they found that radio outbursts tend to repeat every
$\sim8$ years, there is no evidence for flux modulation at the
proposed precession period. 

The non-detection of flux periodicity
at the precession period is not necessarily a serious objection
to the precession hypothesis since a large fraction of the total
radio flux arises from the extended jet which clearly does not
systematically vary in position angle (\citealt*{dmm00}; \citealt*{s03}).

\section{Summary}
We have searched for confirmation of the periodic inner-core
orientation changes reported by \citet{s03} in the parsec-scale
radio jet of the AGN BL Lac.  We analyzed 43~GHz VLBA
observations over ten epochs from 1998.76 -- 2003.82 which
overlap the time range reported by Stirling et al.  As expected,
the resulting maps are very similar in quality and angular
resolution to those of Stirling et al. Using a Gaussian
model-fitting procedure similar to that of \citet{s03} but with
different  criteria for estimating the uncertainties, we find:
\begin{enumerate} 

\item We confirm (at 95\% confidence) variability of the
structural position angle (SPA) of the the inner core. The
acceptance probability of the null hypothesis (no variation) was
$p=0.04,0.01$ for the one and two-component Gaussian model fits
to the inner core.  

\item Our SPA orientation measurements are consistent with 
those of \citet{s03} during overlapping
epochs, after multiplying their uncertainties by a constant factor
of 1.5. This adjustment also makes their periodic model agree (in
a $\chi^2$ statistical sense)
with their SPA data: The acceptance probability is reduced from
$p<10^{-4}$ to $p=0.44$.

\item Our SPA data is best-fit by periodic functions with periods 
12.0 yr ($p=0.38$) and 12.1 yr ($p=0.14$) for one and
two-component Gaussian models respectively. However, the
2.29 yr periodic model of \citeauthor{s03}  was also 
a plausible fit to our data (acceptance probabilities $p=0.18, 0.10$).

\item Combining the rescaled \citet{s03} SPA data with the present dataset, we
find a best-fit periodic model with a period of $2.24\pm0.2$ yrs
($p=0.49$), very close to the 2.29 year period found by
\citeauthor{s03}. 

\item Using the best-fit model, a $\chi^2$ test of agreement with
three SPA measurements made since the last \citet{s03}
epoch (2000.6- 2003.78) results in an acceptance probability
$p=0.08$, i.e. rejection of the periodic model at slightly less than $2\sigma$
confidence.

\end{enumerate} 

If the relativistic jet is precessing, the consequent variation
in line-of-sight angle implies periodic flux modulation of the
inner core component. We estimate the flux ratio to be 1.1 to 2.5
based on jet parameters modeled from studies of superluminal
components (\citealt{dmm00}). However, studies of the radio and
optical light curve of BL Lac over several decades
(\citealt{k03}, \citealt{v04}) find no periodic variation at the
suggested precession period. 

In summary, our although our SPA data provides confirming
evidence for orientation variability in the inner core of BL Lac
consistent with the measurements of \citet{s03}, the evidence for
periodic variations is less compelling. A definitive
resolution will require continued monitoring of the inner core
radio structure over several periods.

\acknowledgements
We thank Casey Dreier for assistance with the data reduction.


\end{document}